\documentclass{PoS}

\newcommand{\sigmaSI}{\sigma_{\rm SI}}
\newcommand{\sigmaSD}{\sigma_{\rm SD}}

\newcommand{\mev}{\rm MeV}
\newcommand{\gev}{{\rm GeV}}
\newcommand{\tev}{\rm TeV}
\newcommand{\pb}{\rm pb}

\newcommand{\eqref}[1]{Eq.~(\ref{#1})}

\newcommand{\lsim}{\lower.7ex\hbox{$\;\stackrel{\textstyle<}{\sim}\;$}}
\newcommand{\gsim}{\lower.7ex\hbox{$\;\stackrel{\textstyle>}{\sim}\;$}}

\newcommand{\be}{\begin{equation}}
\newcommand{\ee}{\end{equation}}

\newcommand{\bea}{\begin{eqnarray}}
\newcommand{\eea}{\end{eqnarray}}

\title{WIMPless Dark Matter: Models and Signatures}

\ShortTitle{WIMPless Dark Matter: Models and Signatures}

\author{\speaker{Jason Kumar}
\\
        University of Hawaii\\
        E-mail: \email{jkumar@hawaii.edu}}

\abstract{We consider experimental signatures of
WIMPless dark matter.  We focus on models where the WIMPless dark
matter candidate is a Majorana fermion, and dark matter scattering is
predominantly spin-dependent.  These models can be probed by IceCube/DeepCore,
which can potentially find $3\sigma$ evidence with $\sim$ 5 years of data.}

\FullConference{35th International Conference of High Energy Physics - ICHEP2010,\\
		July 22-28, 2010\\
		Paris France}

\begin{document}

\section{Introduction}

WIMPless dark matter~\cite{Feng:2008ya} is a robust and versatile scenario in which the dark matter
candidate is a hidden sector particle whose mass is at the hidden sector soft-SUSY breaking
scale.  Although the dark matter candidate can be either a boson or fermion and its mass
can vary over a wide range ($10~\mev \lsim m_X \lsim 10~\tev$), it automatically
has the approximately correct relic density to match cosmological observations.  The versatility
of this scenario implies that specific models can have a wide variety of possible signatures,
detectable using many different dark matter search strategies.

WIMPless dark matter signatures depend on
whether the candidate is a boson or fermion.  If the WIMPless candidate (the lightest
particle stabilized by a hidden sector symmetry) is a scalar,
this WIMPless candidate can have a much large spin-independent scattering cross-section
($\sigmaSI$) than one would expect from neutralino WIMPs.  Models of this type have been
studied~\cite{Feng:2008dz} as a way of explaining the recent data of the DAMA, CoGeNT and
CRESST experiments~\cite{lowmassdata}.  Signatures of these models can be studied at detectors
such as the Tevatron, the LHC, Super-Kamiokande, and liquid
scintillator neutrino detectors~\cite{lowmasstests}.

In these proceedings we focus on a different signature which can be found in a complementary
set of models: spin-dependent scattering.  Several detectors
exist which are sensitive to the spin-dependent scattering cross-section ($\sigmaSD$).  But
bounds on $\sigmaSI$ are much tighter than those on $\sigmaSD$, because spin-independent
scattering receives an $A^2$ enhancement from coherent scattering in the heavy nuclei of a
detector.  Most dark matter models will be probed first by $\sigmaSI$-sensitive
detectors, so it is worth considering which classes of models will probed first by
$\sigmaSD$-sensitive detectors.

An interesting class of such models are WIMPless dark matter models in which the dark matter
candidate is a Majorana fermion.  In these proceedings, we will study models of this type
and consider detection prospects for IceCube/DeepCore.

\section{WIMPless Dark Matter}

WIMPless dark matter is a hidden
sector particle with mass at the hidden sector soft SUSY-breaking scale~\cite{Feng:2008ya}.  The
thermal relic density is set by the annihilation cross-section, $\rho \propto
\langle \sigma_{ann.} v \rangle^{-1}$, which in turn is determined from dimensional
analysis, $\langle \sigma_{ann.} v \rangle \propto {g^4 \over m^2}$.  To get the
correct relic density, one must have $\langle \sigma_{ann.} v \rangle \sim 1~\pb$.
The ``WIMP Miracle" is the remarkable coincidence that, for $g=g_{weak}$ and
$m=m_{weak}$, the annihilation cross-section is indeed $\sim \pb$.  In the
WIMPless scenario, both the MSSM sector and the hidden sector receive the effects of
SUSY-breaking through gauge-mediation from the same SUSY-breaking sector.  The soft
scale is thus set by the gauge-coupling, yielding the relation
\bea
{g^2 \over m} , {g_{weak}^2 \over m_{weak}} \propto {M_{mess.} \over F}
\eea
where $g$ and $m$ are the coupling and soft SUSY-breaking scale of the hidden sector,
$F$ is the SUSY-breaking vev, and $M_{mess.}$ is the messenger mass scale.
The WIMPless dark matter candidate thus naturally has approximately the same
annihilation cross-section (and relic density) as a WIMP.

This is a very robust result, essentially determined by dimensional analysis
and the power-counting of gauge interactions.  The relic density calculation
does not depend on whether the WIMPless candidate is a boson or a fermion.
When the WIMPless candidate is a Majorana fermion, an interesting feature is that
the dark matter-nucleon scattering cross-section may be largely spin-dependent.
WIMPless dark matter can couple to Standard Model matter through Yukawa couplings
\bea
V = \lambda_{Li} \tilde Y_L \bar X P_L f_i + \lambda_{Ri} \tilde Y_R \bar X P_R f_i
+ h.c.
\eea
where the $f_i$ are Standard Model fermions.  Dark matter-nucleon scattering arises through
coupling to Standard Model quarks, and in general WIMPless dark matter can couple
to all generations.

The scattering cross-section can be determined from the Yukawa couplings above.
The spin-dependent part of the scattering cross-section (assuming no squark
mixing) is given by
\bea
\sigmaSD &=&
{m_r^2 \over 4\pi }
{J+1 \over J}
\left[\sum_i \left({|\lambda_{Li}^2| \over m_{\tilde Y_L}^2 - m_X^2 }
+{|\lambda_{Ri}^2| \over m_{\tilde Y_R}^2 - m_X^2 } \right)
\left(\langle S_p \rangle \Delta_i^{(p)} + \langle S_n \rangle \Delta_i^{(n)}\right)
\right]^2\,,\label{MajSD}
\eea
where
\begin{equation}
\Delta_u^{(p)} = \Delta_d^{(n)} = 0.78\pm0.02\,,\,\,\,\,\,\,\,
\Delta_d^{(p)} =\Delta_u^{(n)} = -0.48\pm0.02\,,\,\,\,\,\,\,\,
\Delta_s^{(p,n)} = -0.15\pm0.02\,.\label{Deltas}
\end{equation}
are spin-structure functions.  We will focus on the case where dark matter
couples dominantly to the first generation quarks.
If the WIMPless candidate is a Majorana fermion, then the only diagrams
contributing to spin-independent scattering are those involving mixing of the
4th generation squarks, $\tilde Y_L$ and $\tilde Y_R$.  The spin-independent scattering
cross-section is thus proportional to $Re(\lambda_L \lambda_R^*)^2$.  If squark-mixing
is small, or if $\lambda_L \ll \lambda_R$ (or $\lambda_R \ll \lambda_L$), the dark matter-nucleon
scattering cross-section will be largely spin-dependent.

\section{Tests at IceCube/DeepCore}

\begin{figure}[tb]
\begin{center}
\includegraphics*[width=0.32\columnwidth]{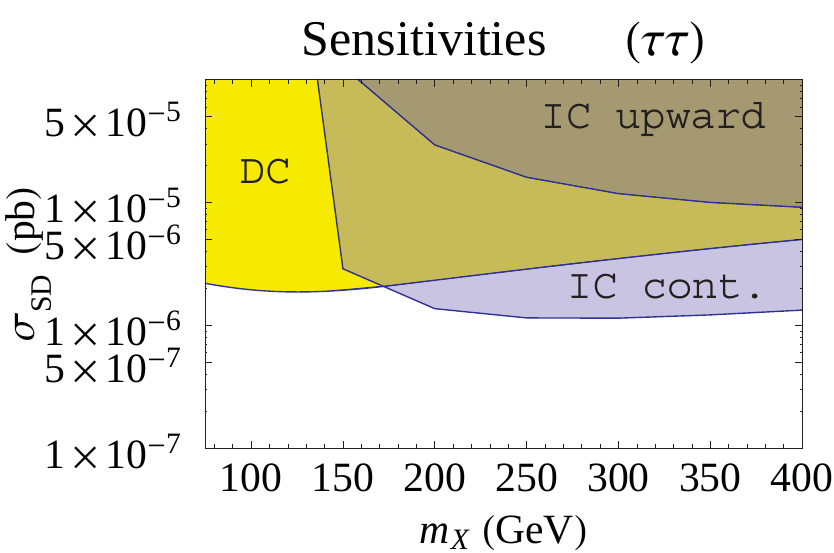}
\includegraphics*[width=0.32\columnwidth]{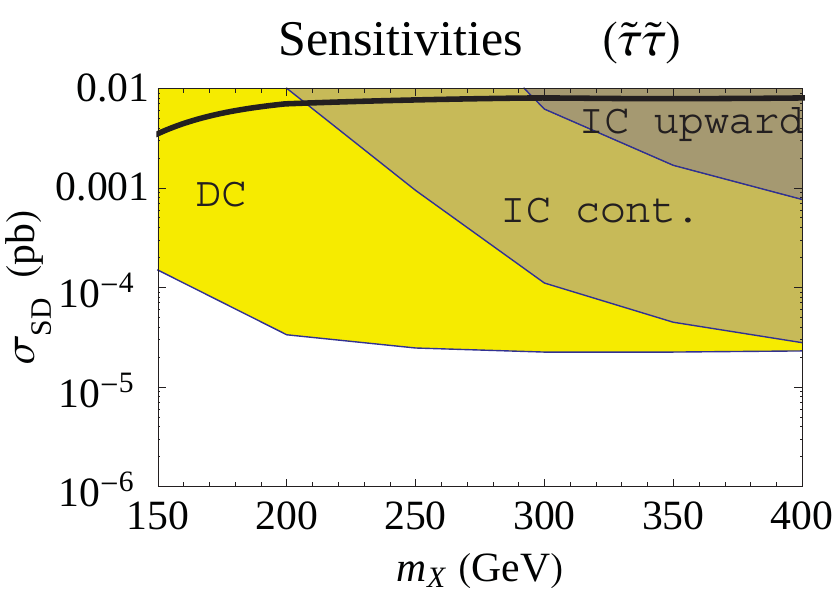}
\includegraphics*[width=0.32\columnwidth]{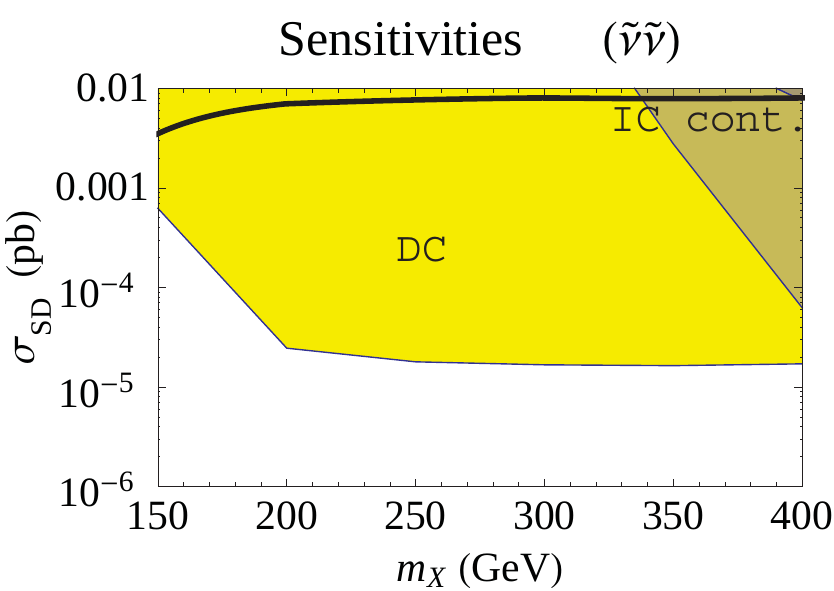}
\end{center}
\vspace*{-.25in}
\caption{\label{fig:tau} 3$\sigma$ detection prospects for Majorana fermion WIMPless
dark matter annihilating exclusively to $\tau \bar \tau$ (left), $\tilde \tau \tilde \tau^*$ (middle)
and $\tilde \nu \tilde \nu^*$ (right).  We assume $m_{\tilde \tau_1} = 137~\gev$,
$m_{\tilde \nu} = 111.5~\gev$, $m_{\tilde \chi_1^0} = 94.5~\gev$, and the decays $\tilde \tau \rightarrow
\tau \tilde \chi_1^0$, $\tilde \nu \rightarrow \nu \tilde \chi_1^0$.  We assume that
the sneutrino Yukawa couplings are flavor-independent.  The dark grey region (IC upward),
blue region (IC contained), and the light brown region (DC) indicate detection prospects using
the upward throughgoing sample from IceCube, the muon sample fully contained
within the IceCube volume, and muon sample fully contained within the DeepCore volume, respectively.
The solid black line is the bound from Super-Kamiokande~\cite{Desai:2004pq}.
(Figure courtesy of Danny Marfatia and Enrico Sessolo.)
}
\end{figure}

The IceCube detector (with the DeepCore extension) is expected to soon have
among the best sensitivity to $\sigmaSD$~\cite{Braun:2009fr}, obtained through bounds on the
neutrino flux from dark matter which annihilates after being captured in the sun.
Detection prospects at IceCube/DeepCore
for such models were studied in~\cite{Barger:2010ng}, including the effects of neutrino propagation
through the sun and vacuum.  Detection prospects for the $\tau \bar \tau$, $\tilde \tau \tilde \tau^*$
and $\tilde \nu \tilde \nu^*$
annihilation channels are shown in Figure~\ref{fig:tau}.
In particular, it was found that $3\sigma$-evidence of models with reasonable Yukawa
couplings ($\lambda_{u,d} \sim 0.5$) could be found with 5 years of running time.  As expected, the
DeepCore extension is most important for low-energy neutrinos, resulting from either low-mass dark
matter annihilation, or from annihilation to superpartners whose decay chain results in lower energy
neutrinos.  For higher mass dark matter, the best detection prospects arise from the events fully-contained
within the IceCube volume.

\section{Conclusion}

WIMPless Majorana fermion models contain a class of well-motivated dark matter
candidates for which IceCube/DeepCore may very well be a discovery experiment.  The relative
detection prospects at IceCube/DeepCore, as opposed to direct detection experiments such
as CDMS or Xenon100, depend on the magnitude of 4th generation squark-mixing, as well as the
relative strengths of the $\lambda_L$ and $\lambda_R$ Yukawa couplings.  A detailed study
of these cross-sections, and their effect on detection prospects, is currently underway~\cite{SIvsSD}.

\acknowledgments

We are grateful to the organizers of ICHEP2010, and to V.~Barger, D.~Marfatia, and
E.~Sessolo for collaboration and J.~L.~Feng, K.~Fukushima and P.~Sandick for discussions.
JK is supported by DOE grant DE-FG02-04ER41291.

\end{document}